\documentclass[12pt]{iopart}
\usepackage[utf8]{inputenc}
\usepackage[T1]{fontenc}
\usepackage{graphicx}
\usepackage[english]{babel}
\usepackage{iopams}
\usepackage{amssymb}
\usepackage{hyperref}
\usepackage{xcolor}
\usepackage{mathptmx}
\usepackage{etoolbox}
\usepackage{siunitx}

\begin{document}
	
\title[Dynamic breaking of axial symmetry in crystals]{Dynamic breaking of axial symmetry of acoustic waves in crystals as the origin of nonlinear elasticity and chaos: Analytical model and MD simulations.}
	
\author{Zbigniew Kozio{\l}}
	
\address{National Center for Nuclear Research, Materials Research Laboratory\\ul. Andrzeja So{\l}tana 7, 05-400 Otwock-{\'S}wierk, Poland}
\ead{zbigniew.koziol@ncbj.gov.pl}

\vspace{10pt}
	
	\begin{abstract}
	 A Chain of Springs and Masses (CSM) model is used in the interpretation of molecular dynamics (MD) simulations of movement of atoms in orientated FCC crystals. A force of dynamic origin is found that is perpendicular to the direction of the external shear pressure. It is proportional to the square of the applied pressure;  It causes breaking of axial symmetry for propagation of transverse acoustic waves. It leads to a non-linear elastic response of crystals and to chaotic patterns in the motion of atoms. We provide an analytical derivation of an effective atomistic 3D potential for interaction between crystallographic layers. The potential is found to possess a component that has an anharmonic threefold axial symmetry around one direction. It reduces to the H{\'e}non-Heinen potential in a 2D cross-section, leading to mathematically rich, complex dynamic features. Results of simulation predict displacements of atoms that are inconsistent with the static theory of elasticity that may have been overlooked in experiments. 
	\end{abstract}

\vspace{2pc}
	
	\maketitle

\section{Introduction.}
\label{Introduction.}

\subsection{The model of a Chain of Springs and Masses (CSM).}
\label{CSM}

The CSM model, based on the works of Schrödinger (1914) \cite{Erwin,Muhlich} and Pater (1974)\cite{Pater}, has been found recently \cite{Stretched} suitable as an analytical description of the dynamics of crystallographic layers in orientated FCC crystals studied by the Molecular Dynamics (MD) technique. On the side of MD simulations, the method is to determine the time evolution of averages of some physical quantities when an external pressure is applied to the outer surface of the sample. These quantities 
are, for instance, $\langle X_{i,n}\rangle$, $\langle V_{i,n}\rangle$ (displacement and velocity of atoms, respectively, in arbitrary direction $i$). $\langle\rangle$ means an average over all atoms within the plane, while index n enumerates planes, starting with n = 1 for the surface rigid plane, where force is applied, and with n = N for the plane with the lowest Y value, which is fixed (planes are perpendicular to the Y-direction, which is the [111] crystallographic direction, while external force is applied in the X-direction [1$\bar{1}$0], and the Z-axis is along [$\bar{1}$$\bar{1}$2]). That geometry is typical in simulations where studies of the dynamics of dislocations are performed in FCC crystals \cite{Osetsky}. The symmetry conditions imposed in the \texttt{lammps}\cite{LAMMPS} simulation tool are written as \texttt{psp}, which means that atoms can move and interact with any atoms through the simulation box boundary in directions X and Z, but not in direction Y. This is apparently the main reason why a collection of atoms behaves like it was forming a chain of layers (with nearly rigidly connected atoms within each layer), while layers interact effectively through a spring between them. There is evidence of the validity of the CSM model based on numerous simulations, of which a part only has been published \cite{Stretched} so far.

The model predicts some surprising, mathematically rich features. For instance, the sound speed is found to be dependent on the time/distance of the wave travelled, and it depends also on how the threshold value of the sound wave is defined in any experiments. That is caused by a change of sound wave profile with time/distance. Oscillations in quantities like velocity or forces were found, and their frequency changes with time. Amplitudes of oscillations have a time-dependence close to a power-law or a stretched-exponential function of time.

\subsection{The motivation.}
\label{motvation}

It follows from elementary symmetry considerations that Cauchy forces in the Z-direction (and in the Y-direction as well) ought to be equal to zero when a static shear pressure is applied in the X-direction at the surface. That implies also the absence of any movement of atoms (layers) in Z- and Y-directions.

However, in our simulations (which are, obviously, carried out in a dynamic way) we observe notoriously that some small forces (a few orders of magnitude smaller than forces in the X-direction) are observed. These forces are proportional to the square of the applied pressure in the X-direction, and, therefore, they diminish quickly when the pressure value is very small.

Moreover, we are able to determine the dynamics in the Z-direction, finding out that there is no exact agreement between forces and acceleration as it ought to follow from Newtonian mechanics. However, if a small contribution to forces in Z is added from forces in X-, such an agreement could be obtained.
In other words, we observe a mixing of X- and Z-components of forces (movement), which contradicts the common belief of what ought to be observed in a case of angularly isotropic potential in a plane (1,1,1).
 
The initial aim of this work was to gain an understanding of the mechanisms behind the occurrence of frequency beats in oscillations observed at large pressures applied in quantities like the velocity of layers, or the force acting on them, as reported first in \cite{Stretched}; see figure 6 there. However, we will not discuss here these frequency beats, leaving the subject for another article. Instead, we will concentrate on finding out a mechanism of erasing of an unusual force component contributing to the dynamics of stress. Actually, that force component of a dynamical origin - to the best of our knowledge, not discussed so far - is the cause of precession in 2D phase space, leading to frequency beats.

Yet another important finding, though not planned by us, follows from the analysis of the 3D equation on the potential energy. The derived
function $E_p(x,y,z)$ is an approximation obtained by taking the first terms of the Taylor series expansion of an exact potential. It is found 
as a sum of axially symmetric harmonic contributions in $x$ and $z$, another harmonic contribution in $y$, and a term that has a three-fold axial symmetry around $y$. For $y$=0, the potential energy reduces to a form given by the famous H{\'e}non-Hiles equation (1964) \cite{Henon} inspired by analysis of the motion of stars in the Galaxy that sparked the development of the theory of chaos.

\subsection{Samples, potentials and simulation setup.}
\label{samplesPotentials}

We use as simple as possible crystallographic structure and interatomic potential that allows us to reproduce the most basic dynamical mechanisms similar to those in real systems, where line dislocation movement is observed in an orientated FCC lattice. A harmonic interaction between atoms is assumed and a perfectly ordered monoatomic crystal. The details of the sample preparation, geometric configuration, creating the interatomic potential, MD simulations, and the data analysis methods have been described previously \cite{Stretched,anharmonic}. 

This time we use the \texttt{lepton} interatomic potential available in \texttt{lammps}, alongside the previously used by us - the \texttt{table} style potential of our design \cite{harmonic} - in experiments described here, both potentials lead to identical results. 

A proper statistical averaging of physical quantities along the X direction, for every layer of atoms in the Y-direction is one of the most important steps in the data analysis, and it is probably causing the largest difficulty to other researchers attempting to repeat similar simulations. Some related technical details are given in \cite{averaging}.

In simulations of the long-time dynamics of the CSM model, we the inter-atomic potential $E_p(r)=\epsilon_0\cdot (r-r_0)^2$,
with the parameter $\epsilon_0$ = 1 eV/{\AA}$^2$, where $r$ is the distance between atoms, $r_0=a/\sqrt{2}$, and $a$ is the FCC lattice constant, the same as in steel, $a$ = 3.56 {\AA}, with the mass of atoms assumed as that of Fe.

Results reported in \cite{Stretched} were obtained, mostly, on a sample with 8 atoms in one layer. We, however, realised (and checked carefully) that only two atoms in a layer are sufficient for carrying out most of the MD simulations described here (that speeds up the simulation time, data analysis time and decreases the amount of stored data). There are some exceptions to that when it was necessary to use larger sample sizes. Here, the sample has N=1500 layers in the Y-direction (The length is over 0.3 $\mu$m).

The results of simulations are, usually, very accurate, exceeding 5-6 significant digits.

\subsection{The plan of this article.}
\label{ThePlan}

In section \ref{XZdynamics} we compare how the time dependencies of certain physical quantities (displacement, velocity, force) differ for directions X and Z when the free movement of layers/particles/atoms in the Z-direction is enabled. We show there a few phase-space diagrams, bringing some light on the possible mechanisms behind the interaction between particles.

In section \ref{OriginZ-forces}, which is the crucial one, we provide equations on the interlayer potential depending on coordinates x, y, and z. We show that it has the main harmonic contribution that is axially symmetric and a small contribution that grows up as a third power of distance from its minimum and has a threefold axial symmetry. We provide evidence based on MD simulation results of, e.g., angular dependence of forces, that the simulations are in agreement with the derived potential energy. We also propose approximate equations of motion of a particle at short times and show how the dynamics of the system lead to chaotic motion.

In discussion section \ref{discussion} we point out how the proposed description relates to the Christoffel equations that describe atomic motions by using a plane-wave linear formalism, and we discuss briefly a connection between the observed dynamics of layers and the H{\'e}non-Heiles model. Finally, we attempt to formulate how to solve the problem of the motion of waves in the bulk of material by using a pair of nonlinear coupled difference equations. Possible experimental consequences of the model are mentioned as well.

\section{Comparison of dynamics in X- and Z-directions.}
\label{XZdynamics}

\subsection{The linear case, with no movement in the Z-direction.}

At low values of the applied surface pressure, let's say below 10 MPa, the results of MD simulations can be described exactly by solutions of a 1-dimensional CSM model, as shown in \cite{Stretched}. Hence, after an abrupt pressure is applied at the surface (Heaviside mode of simulations) the time-dependent quantities like the displacement of layers, ${u}_{n}^{H}$, their velocity ${v}_{n}^{H}$ and the force acting on them, ${f}_{n}^{H}$, has the form:

\begin{eqnarray}&&{u}_{n}^{H}=\frac{{F}_{1}}{{\rm{m}}{{\rm{\Omega }}}^{2}}\cdot \left[{J}_{2n}(2\theta )+\displaystyle \sum _{k=2n+2,2n+4,...}^{\infty }(k-2n+1){J}_{k}(2\theta )\right],
\label{eq:uH}
\end{eqnarray}

\begin{eqnarray}&&{v}_{n}^{H}=\frac{{F}_{1}}{{\rm{m}}{\rm{\Omega }}}\cdot \left[{J}_{2n-1}(2\theta )+2\cdot \displaystyle \sum _{k=2n+1,2n+3,...}^{\infty }{J}_{k}(2\theta )\right],
\label{eq:vH}
\end{eqnarray}

\begin{eqnarray}&&{f}_{n}^{H}=-{F}_{1}\cdot \left[{J}_{2n}(2\theta )+2\cdot \displaystyle \sum _{k=2n+2,2n+4,...}^{\infty}{J}_{k}(2\theta )\right],
\label{eq:fH}
\end{eqnarray}

where $J_n$ are Bessel functions of the first kind and $\theta = \Omega t$, \rm{m} is the mass of a particle/layer, and $F_1$ is the force exerted on the uppermost layer. The parameter $\Omega$ used there is the fundamental angular frequency of oscillations of two layers: it can be determined by using several methods from MD simulations, and it is in agreement with the value computed from first principles \cite{Stretched}. The superscript $^H$ refers to the Heaviside method of simulations.

A very similar, numerically nearly identical one is a solution for an often-used case when a constant rate of the surface-applied strain is used in MD simulations. Let us list these equations as well, for a reference (the superscript $^V$ is used to distinguish the case from the Heaviside mode of simulations):

\begin{equation}
u^V_n(t) = \frac{v_0}{\Omega } \left[J_{2n-1}(2\theta) + \sum _{k=2n+1,2n+3,...} ^{\infty} (k-2n+2) J_k(2\theta)\right],
\label{eq:Vun}
\end{equation}

\begin{equation}
v^V_n(t) = v_0\left[J_{2n-2}(2\theta) + 2\cdot \sum _{k=2n,2n+2,...} ^{\infty} J_k(2\theta)\right],
\label{eq:Vvn}
\end{equation}

\begin{equation}
f^V_n(t) = -m v_0 \Omega \cdot \left[ J_{2n-1}(2\theta) +2\cdot \sum _{k=2n+1,2n+3,...} ^{\infty} J_k(2\theta)\right].
\label{eq:Vfn}
\end{equation}

In \ref{eq:Vun}-\ref{eq:Vfn}, $v_0$ is the imposed constant velocity of the surface layer. These equations have been deduced from accurate analysis of results of MD simulations. In the constant surface velocity method (which is equivalent to the constant surface strain rate $\dot{\epsilon}=v_0/d$, where $d$ is the sample size in the direction of the applied force), the equations are related to these in the Heaviside method, with velocity $v_0$ in \ref{eq:Vvn} having the same value as the prefactor in \ref{eq:vH}, with the force at the surface, $F_{1}$, creating the same pressure values in the sample volume of the constant velocity mode. 

In the linear case, as given by \ref{eq:uH}-\ref{eq:fH}, simulations do not show, obviously, any observable contribution to quantities in the Z-direction, i.e., in the displacement, velocity, or force, $\Delta Z$, $V_Z$ or $F_Z$. 

The linear case is realised in simulations by two methods: \textbf{1}) when the applied pressure is very low, the system response may be considered linear as a function of pressure, or \textbf{2}) when a condition is imposed in \texttt{lammps} that forces do not act on atoms in the Z-direction, by a command like this: \\
\texttt{fix~id~all~setforce~NULL~0~0}. 

In the first method any response scales up linearly with the applied pressure. The pressure range where this is valid is up to around 10 MPa (which is a somewhat arbitrary value, depending on the desired accuracy). In case \textbf{2}) the linear response is preserved up to around 10 GPa (i.e., up to the largest pressure values we can apply).

\subsection{The nonlinear case, with movement in the Z-direction enabled.}
\label{Zallowed}

\begin{figure}[ht]
\centering
\includegraphics[scale=1.0]{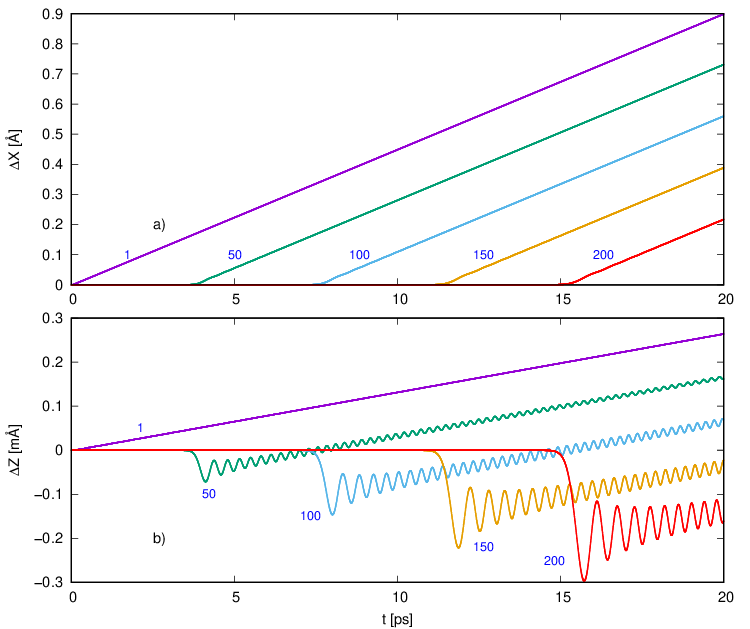}
\caption{
Comparison of average displacement of layers in the X-direction (figure a) and in the Z-direction (b),
when the shear pressure of 100 MPa is applied at the surface in the X-direction.
Labels indicate the number of layer. 
}\label{fig:XYZL00A2}
\end{figure}

\begin{figure}[ht]
\centering
\includegraphics[scale=1.0]{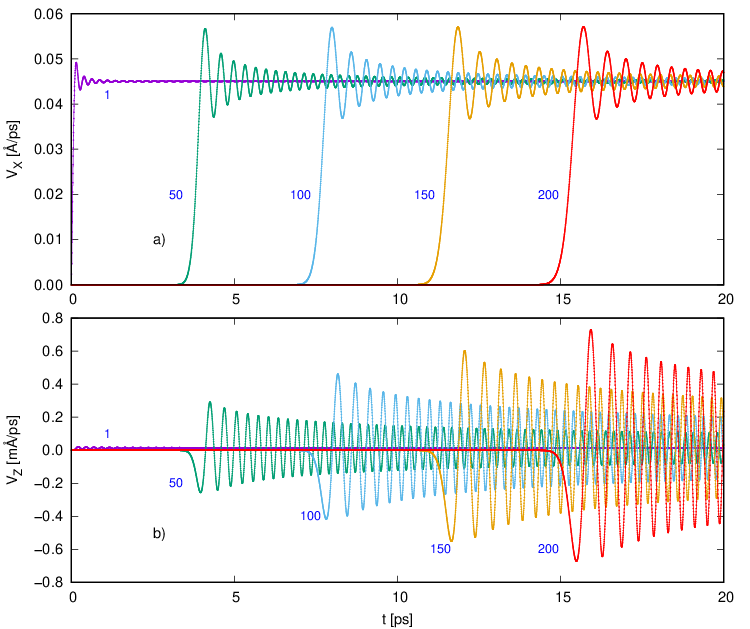}
\caption{
Comparison of average velocity of layers in the X- and Z-directions, $V_X$ and $V_Z$, under the same simulation conditions and for the same layers as in figure \ref{fig:XYZL00A2}. Velocities are time derivatives of displacement curves shown in figure \ref{fig:XYZL00A2}. 
}\label{fig:XYZL00D2}
\end{figure}

\begin{figure}[ht]
\centering
\includegraphics[scale=1.0]{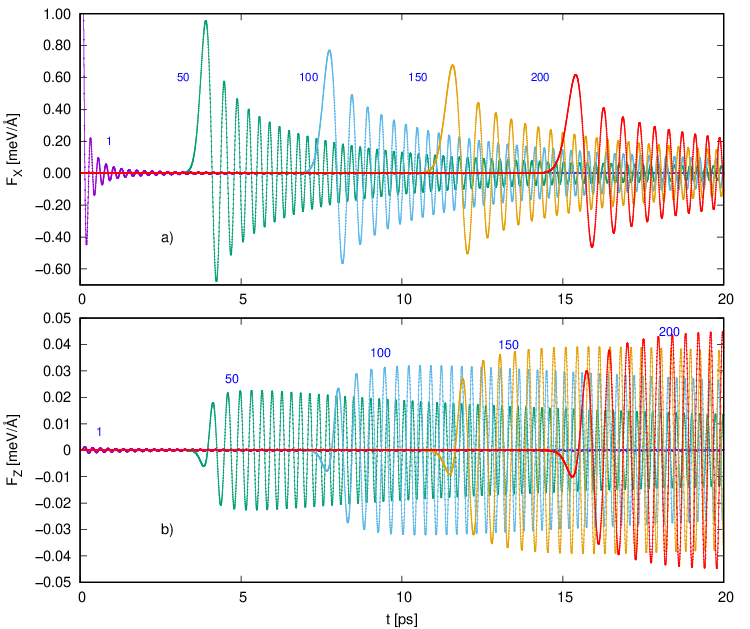}
\caption{
Comparison of average forces acting on layers in the X- and Z-directions, under the same simulation conditions and for the same layers as in figures \ref{fig:XYZL00A2} and \ref{fig:XYZL00D2}. Notice the difference in vertical scale in a) and b).
}\label{fig:XYZL00E2}
\end{figure}

Figures \ref{fig:XYZL00A2}-\ref{fig:XYZL00E2} show a comparison of some physical quantities in directions X and Z when atoms are free to move in the X and Z directions, and the system is in the nonlinear regime after we use \\
\texttt{fix~id~all~setforce~NULL~0~NULL}.

We do not show the results on quantities in the Y-direction when the movement in that direction is allowed in order to avoid complicating this description. Those are basically similar to these in the Z-direction.

At a pressure of 100 MPa, as in figures \ref{fig:XYZL00A2}-\ref{fig:XYZL00E2}, the X quantities are still responding nearly linearly to pressure, while Z quantities are small but already well visible in simulations and scale up as the second power of pressure. Notice the difference in vertical scales between a) and b) in figures \ref{fig:XYZL00A2}-\ref{fig:XYZL00E2}.

The slope dX/dt (after the pressure wave reached the layer) is 0.04502 {\AA/ps}. It is an average velocity of layers in the X-direction, $V_X$. When performing simulations with disabled movement in the Z-direction, that velocity is proportional to the applied pressure, with a high accuracy, from the lowest pressure values investigated (of the order of 10$^{-3}$ MPa) up to around 8 GPa: $V_X = 0.0004502\cdot P$ {\AA/ps}, where $P$ is in MPa. The proportionality coefficient is in perfect agreement with the value computed by using the CSM model \cite{Stretched}. 

It was unexpected to observe, as in figure \ref{fig:XYZL00A2} b), that a displacement in the Z-direction is present as well. Moreover, when simulations are done by applying force in the opposite direction (still, along the X-axis) we observe an increase of Z(t) as well, following exactly the same curves, showing that the sign of the  displacement Z(t) does not depend on the sign of the applied force. 
Additionally, we could compensate Z(t) to zero value by adding in simulations a very small force component in the Z-direction.

The slope of dZ/dt =0.001323 {\AA}/ps in b) is the same for all layers. 

dZ/dt is the average velocity of layers in the Z-direction, $V_Z$. That velocity has a different dependence on the applied pressure than $V_X$. With a high accuracy, it can be approximated by a parabolic function of pressure, 
$V_Z \sim P^2$. 

The sign of velocity $V_X$ depends on the sign of the applied pressure.

In the case of $V_X$ in \ref{fig:XYZL00D2} a), we can see clearly that at long times velocity oscillates around a certain equilibrium value; we name it $V_{x0}$. The same occurs in the case of $V_Z$, i.e., at an asymptotically long time a value $V_{z0}$ is reached, though in the scale of \ref{fig:XYZL00D2} b) that cannot be seen, because that value is very small.

\subsection{Some phase-space diagrams.}
\label{PhaseSpace}

Before drawing some hypotheses about the mechanisms behind the observed dependencies, let us analyse a few 
phase-space diagrams. 

\begin{figure}[ht]
\centering
\includegraphics[scale=1.0]{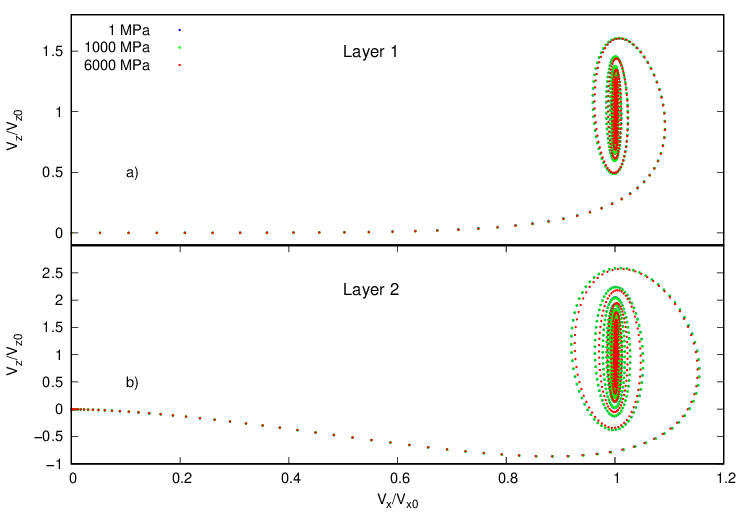}
\caption{Phase-space portrait of $V_z(V_x)$ for three values of pressure, as shown in the legend, when pressure is applied at an angle $\ang{0}$ with respect to the X-axis. a) shows dependencies for layer 1 and b) for layer 2). $V_{x0}$ scales up linearly with pressure; therefore, its value changes here by 6000 times, while $V_{z0}$ changes as $P^2$; hence, it changes here $36\cdot 10^6$ times. 
}\label{fig:AngularV10} 
\end{figure}

In figure \ref{fig:AngularV10} the dependence of $V_z(V_x)$ is shown, where both quantities are normalised by their respective values at asymptotically long times, i.e., by $V_{z0}$ and $V_{x0}$.
As we see, the layer 1 moves initially, at short times along the direction of the applied force, while the movement of layer 2 deviates strongly from the direction of surface force right from $t$=0. 
These results suggest that there is some difference between the dynamic behaviour of the first and the next layers.

\begin{figure}[ht]
\centering
\includegraphics[scale=1.0]{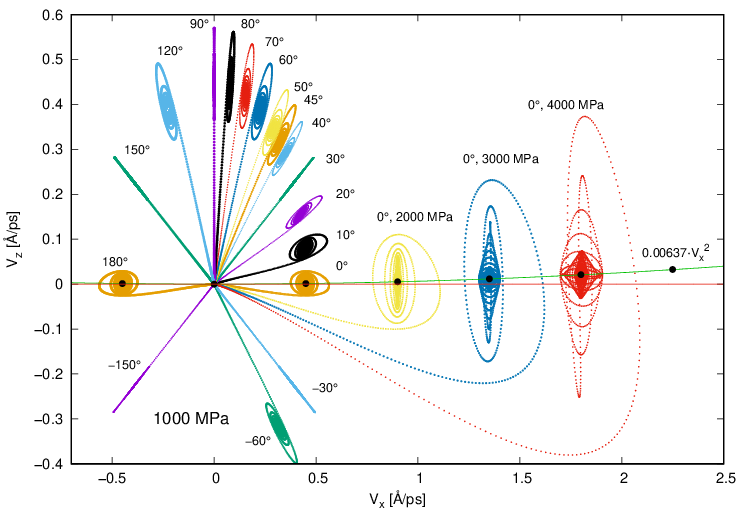}
\caption{
Phase space dependencies $V_{z}(V_{x})$ for layer 50 (time is a hidden parameter there). For a pressure of 1000 MPa a number of curves are shown for a set of angle values, where angles are the direction of the applied pressure at the surface. For pressure applied at an angle of $\ang{0}$ with respect to the X-axis, there are also curves drawn for 2000, 3000 and 4000 MPa. 
}\label{fig:AngularV01A} 
\end{figure}

The results shown so far suggest the existence of a force in the Z-direction when the surface force is applied in the X-direction. Hence, we conclude that in general, the results may depend on the direction of the applied force. Therefore, we carried out simulations for different angles $\alpha$ of the applied force with respect to the X-axis. This is done by defining in \texttt{lammps} the force (pressure) components by a command like this: \\
\texttt{fix id upper aveforce Fx 0 Fz},\\
where $Fx=F1\cdot \cos(\alpha)$ and $Fz=F1\cdot \sin(\alpha)$ for a force of amplitude $F1$.
Figure \ref{fig:AngularV01A} shows the results for $V_{z}(V_{x})$ for several angles $\alpha$ when the pressure has a value of 1000 MPa, and also for 2000, 3000 and 4000 MPa when $\alpha=\ang{0}$.
We would expect that these curves are simply straight lines at the same angles with respect to the X-axis as the applied external force. However, this is not the case. Only for angles of $\ang{30}$ $\pm$ integer multiples of $\ang{60}$ the dependence $V_{z}(V_{x})$ is linear. 
In the limit of long times, well-defined values of velocities for $V_x$ and $V_z$ are observed, $V_{x0}$ and $V_{z0}$, respectively. These are marked, for the angle $\ang{0}$, by black dots on a green curve given by a parabolic function. A linear dependence on pressure is observed, $V_{x0} = 4.502~ 10^{-4} \cdot P$, where P is in MPa and $V_{x0}$ is in {\AA}/ps. That dependence is well understood, and it follows from exact analytical solutions of equations of motion, as described in \cite{Stretched}. For $V_{z0}(P)$ the dependence is of a parabolic type: 
$V_{z0} = 1.29~10^{-9} \cdot P^2$. Hence, that results in the observed dependence $V_{z0} = 0.00637\cdot V_{x0}^2$, as shown by the green line.

\section{On the origin of forces.}
\label{OriginZ-forces}

\subsection{The interlayer potential.}
\label{InterlayerPotential}

\begin{figure}[ht]
\centering
\includegraphics[scale=1.0]{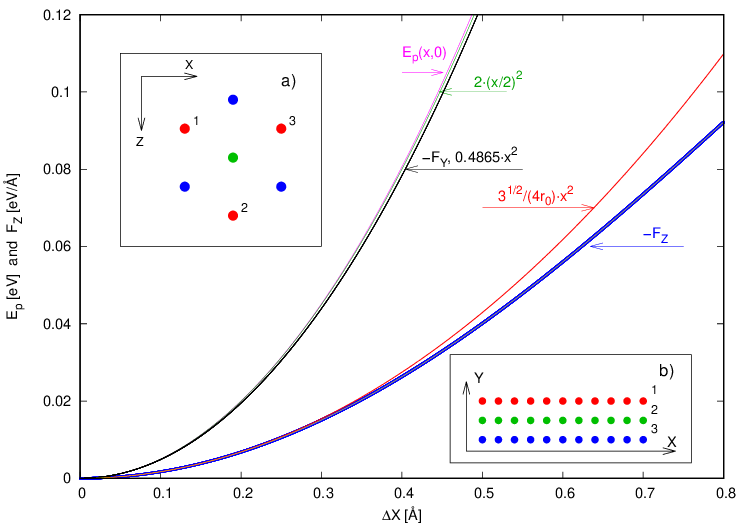}
\caption{
Potential energy $E_p$ and force components $F_Z$ and $F_Y$ for the uppermost layer have been determined from MD simulations
when layer 1 is moved in the X-direction while layers 2 and 3 do not move. 
$E_p(x,0)$ (magenta curve) is the potential energy $E_p(x,z)$ computed by using Eq. \ref{eq:Exz}, when $z$=0. The green curve shows its approximation at low x-values by a parabolic dependence. 
Densely packed blue dots show an $-F_Z$ component of force, as determined from simulations, with the red curve being its approximation by a parabolic dependence, while the tiny yellow line drawn on top of blue data points is computed numerically using \ref{eq:Exz}.
Inset boxes a) and b) show schematic arrangements of atoms during simulations, as described in the main text.
}
\label{fig:myPotential}
\end{figure}

The insets of figure \ref{fig:myPotential}, a) and b), show the arrangement of atoms in layers 1,2, and 3, where layer 1 is the one where force is applied. That arrangement is shown for two different orientations. Colours of atoms in a) are the same as in b): The uppermost layer 1 is coloured in red, the second layer is in green (only one atom is shown in a) in that case), and the lowest layer is in blue. Hence, in a) the green atom has 3 NN atoms in the upper layer in red and 3 NN atoms in the lower layer in blue.

In \cite{Stretched}, when computing the potential energy change due to a displacement of the green atom in the X-direction, we assumed that the arrangement of atoms in the upper and lower layers (red and blue, respectively) does not change. In that case an isotropic potential in the XZ plane is obtained.

However, in our dynamic simulations the wave of stress propagates along the Y-direction. That leads first to a displacement in the X-direction of the upper layer. After a time needed for passing the distance between 2 layers by the propagating wave of pressure, the force acts on the lower layer. In other words, there is some time difference between the displacement of layers 1 and 2. That causes a difference, at any instant of time, between forces acting on layer 2 from the side of layers 1 and 3.

Before proceeding to simulation experiments, let us first discuss equations on the potential energy when a displacement in $x$, $z$, and $y$ is taken into account, in a full analogy to what is done in \cite{Stretched}.

The distance between the green atom in \ref{fig:myPotential} a) and the 3 nearest neighbour (red) atoms on the next plane can be written as

\begin{equation} r_1\left(x,y,z\right) = r_0\cdot \sqrt{1 -\frac{x}{r_0} + \left(\frac{x}{r_0}\right)^2  
+\frac{2\sqrt{2}y}{\sqrt{3}r_0} + \left(\frac{y}{r_0}\right)^2
-\frac{z}{\sqrt{3}r_0} + \left(\frac{z}{r_0}\right)^2}, \end{equation}

\begin{equation} r_2\left(x,y,z\right) = r_0\cdot \sqrt{ 1 + \left(\frac{x}{r_0}\right)^2  
+\frac{2\sqrt{2}y}{\sqrt{3}r_0} + \left(\frac{y}{r_0}\right)^2
+\frac{2z}{\sqrt{3}r_0} + \left(\frac{z}{r_0}\right)^2}, \end{equation}

\begin{equation} r_3\left(x,y,z\right) = r_0\cdot \sqrt{1 +\frac{x}{r_0} + \left(\frac{x}{r_0}\right)^2  
+\frac{2\sqrt{2}y}{\sqrt{3}r_0} + \left(\frac{y}{r_0}\right)^2
-\frac{z}{\sqrt{3}r_0} + \left(\frac{z}{r_0}\right)^2}. \end{equation}

The subscripts in $r_1$,$r_2$,$r_3$ refer to labels of atoms in \ref{fig:myPotential} a). 

The potential energy $E_p(x,y,z)$ of an atom can be written therefore as

\begin{equation} 
	E_p(x,y,z)=\epsilon_0\left[  \left(r_1(x,y,z)-r_0\right)^{2} + \left(r_2(x,y,z)-r_0\right)^{2} +\left(r_3(x,y,z)-r_0\right)^{2}   \right].
	\label{eq:Exz}
\end{equation} 

\begin{figure}[ht]
\centering
$\begin{array}{cc}
\includegraphics[scale=0.65]{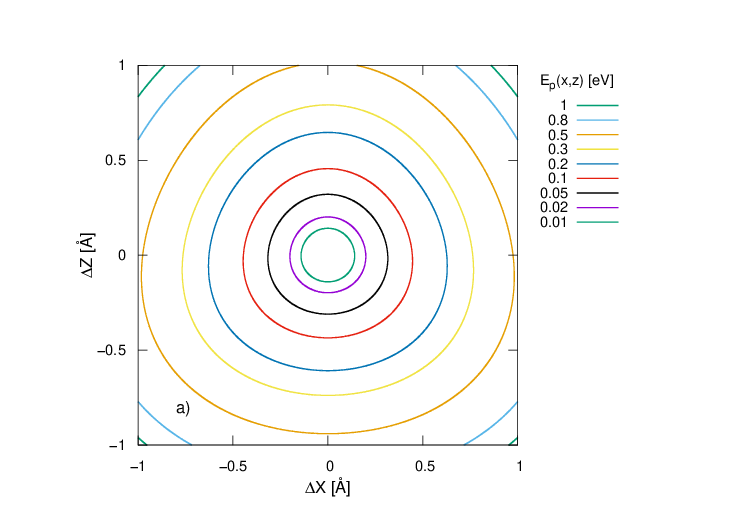}&\includegraphics[scale=0.65]{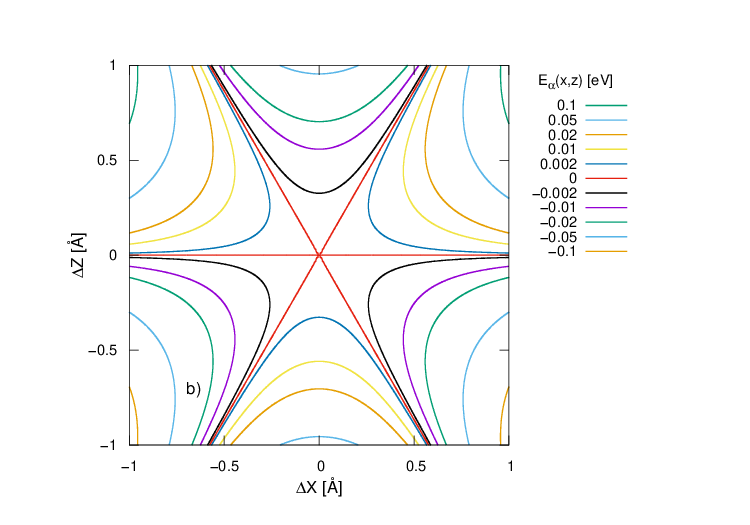}
\end{array}
$\caption{Iso-potential energy countors: a) computed by using an exact equation \ref{eq:Exz} for the total potential energy. b) shows the angular contribution to $E_p$, as given by \ref{eq:D114}.
}
\label{fig:myPotentialA}
\end{figure}

We used equation \ref{eq:Exz} to compute numerically the contours of iso-potential curves for the $x-z$ plane when $y$=0, as shown in \ref{fig:myPotentialA} a). 

One can show, similarly as done in \cite{Stretched}, that to the lowest order of $x$, $y$, $z$, the potential as given by Eq. \ref{eq:Exz} is axially symmetric for rotations around the Y-axis, and it can be approximated by:

\begin{equation} E_p\left(x,y,z\right) \approx 2\epsilon_0 \cdot \left(\left(\frac{x}{2}\right)^2
+4\cdot \left(\frac{y}{2}\right)^2+\left(\frac{z}{2}\right)^2\right). 	\label{eq:Exz2}
\end{equation}

However, the iso-potential curves in \ref{fig:myPotentialA} a) deviate strongly from circles. Moreover, as it will become clear, a harmonic potential cannot explain effects we are going to discuss. Therefore, it become necessary to derive a better approximation of $E_p(x,y,z)$.
A reasonably good one is obtained when a Taylor series expansion of the potential given by \ref{eq:Exz} is done up to the third order in variables $x$, $y$, $z$. However, computing it occurred as a challenging task due to complexity of resulting equations and a large length of the full derivation. For these reasons, we decided to publish the derivation separately \cite{Notes}. There, one can find a derivation up to the 5th order as well, but the resulting equations are significantly more complex. The third order expansion provides an approximation that describes well most of the features discussed here, provided that we limit its application to displacements not larger than around 0.2 {\AA}. The 5th order expansion works up to around 0.5 {\AA}. 

The 3rd order equation can be written in an astonishingly simple form, as a sum of three terms, $E_h$, $E_{\alpha}$, and $E_Y$, i.e., a harmonic one, an angular one, and a term that lacks Y-reflection symmetry, respectively:

\begin{equation}
E_p/\epsilon_0   = \frac{1}{2} \left(x^2+z^2+ 4 y^2\right) 
	+\lambda \cdot\left(x^2 z - \frac{z^3}{3}\right)
	+ \lambda_Y \cdot \left( (x^2+z^2) \cdot y +\frac{2}{3}\cdot y^3 \right),
	\label{eq:Final00xyz}
\end{equation} 

where $\lambda=\sqrt{3}/(4r_0) \approx 0.1720$ {\AA}$^{-1}$ in our case, and $\lambda_Y=\sqrt{3}/(\sqrt{2}r_0)= 2\sqrt{2}\lambda$.

The angular contribution to $E_p$ is given by the term with $\lambda$ in \ref{eq:Final00xyz}. Using the relations between $(x,z)$ in Cartesian and in cylindrical coordinates, $z=\rho\cdot \sin\alpha$, $x=\rho\cdot \cos\alpha$, $\rho=(x^2+z^2)^{1/2}$, as well 
an easy verifiable relation $\sin(3\alpha) = 3 \cos^2\alpha \sin\alpha - \sin^3\alpha$, we can write

\begin{equation} 
E_{\alpha}/\epsilon_0 = \lambda\cdot \rho^3  \cdot \sin(3\alpha)/3.
	\label{eq:D114}
\end{equation} 

That contribution to $E_p$ is shown in \ref{fig:myPotentialA} b).

\subsection{Z-forces.}
\label{Z-forces}

Now, we may compute forces based on derivatives of potential, $F_i = -\partial E_p/\partial x_i$. The force $F_Z(x)$ shown in figure 
\ref{fig:myPotential} is given as a partial derivative of \ref{eq:Final00xyz} over $z$, when $y=0$ and $z=0$:
				
\begin{equation} 
F_Z = -\frac{\partial E_p(x,y,z)}{\partial z}\Big|_{y=0,z=0} = -\epsilon_0\cdot \lambda \cdot x^2.
	\label{eq:Exz3}
\end{equation}

Hence, the parabolic term in \ref{eq:Exz3} is in agreement with the curvature determined in simulations (a derivative computed numerically by using \ref{eq:Exz} is in perfect agreement with simulated results as well, up to $x$=0.8 {\AA}, as shown in \ref{fig:myPotential}).

Encouraged by the observed agreement shown above, we decided to investigate what is the angular dependence of forces that are perpendicular to
the direction of the applied pressure.

\subsection{Angular dependence of forces.}
\label{AngularVelocities}

\begin{figure}[ht]
\centering
\includegraphics[scale=1.0]{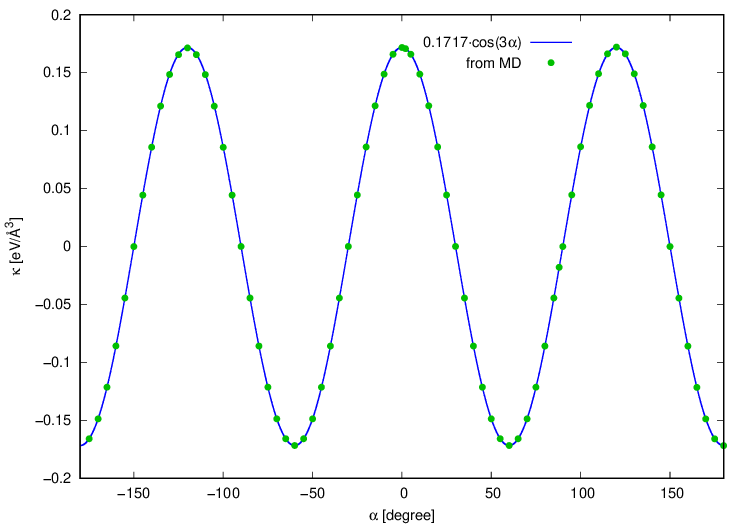}
\caption{
Angular dependence of $\kappa$.
Blue dots labelled as "from MD" have been obtained by fitting parabolic dependence to $F_{\alpha}(r)$, as described in the text. These data are very well described by a function $\kappa=\kappa_0\cdot \cos(3\alpha)$, with $\kappa_0$=0.1717 eV/{\AA}$^3$, as shown by the blue line. }
\label{fig:surfaceXZ1}
\end{figure}

Figure \ref{fig:myPotential} shows the dependence of $F_Z$ on a displacement in the X-direction, i.e., when the displacement direction has an angle of $\ang{0}$ with respect to the X-axis. In order to investigate the angular dependence of potential and forces, we performed similar simulations for an arbitrary angle of the displacement with respect to the X-axis. 

For that, we registered average forces $F_X$ and $F_Z$ as a function of the displacement distance $\rho$, for a large number of angles $\alpha$. In these cases, the displacement vector is given by $\vec{\rho}=\rho\cdot \hat{\rho}$, 
where $\hat{\rho}$ is a unit vector, $\hat{\rho}=(\cos\alpha, \sin \alpha)$. The aim was to find out a force in the direction $\hat{\alpha}$, which is perpendicular to $\hat{\rho}$. From the condition or orthogonality between these unit vectors, $\hat{\rho} \cdot \hat{\alpha} =0$, we find out that $\hat{\alpha}$ is defined as: $\hat{\alpha} =(-\sin\alpha,\cos\alpha)$. Therefore, when the force $F$ is given as $\vec{F}=(F_X,F_Z)$, the force acting in the $\hat{\alpha}$ direction must be given by

\begin{equation} 
F_{\alpha}= \hat{\alpha} \cdot \vec{F} = -\sin\alpha \cdot F_X + \cos\alpha \cdot F_Z.  	
\label{eq:Falpha}
\end{equation} 

After performing simulations in direction $\hat{\alpha}$, quantity $F_{\alpha}$ as given by \ref{eq:Falpha} has been drawn as a function of the displacement $\rho=(x^2+z^2)^{1/2}$, and the dependence 
$F_{\alpha}(\rho)$ was analysed to find out the parabolic coefficient $\kappa$ defined as in the relation 
$F_{\alpha}(\rho) = \kappa \cdot \rho^2$. All these dependencies are similar to that one shown for $F_Z$ in figure \ref{fig:myPotential}, with a strong angular dependence of $\kappa$ found (figure \ref{fig:surfaceXZ1}), given by the function:

\begin{equation} 
\kappa=\kappa_0\cdot \cos(3\alpha).
\label{eq:kappa}
\end{equation} 

When a parabolic approximation is used to find $\kappa_Y$ assuming $F_{Y}=\kappa_Y\cdot \rho^2$,
for the displacement $\rho$ up to 0.1 {\AA}, an average value of $\kappa_Y= -0.4867$ eV/{\AA}$^3$ is found, which is very close to -0.4865 eV/{\AA}$^3$ predicted by \ref{eq:Final00xyz}. There is a 1.5\% angular contribution of $\sin(3\alpha)$ there as well, which we will ignore.

\subsection{Displacement experiments.}
\label{Trajectories}

In this subsection we propose yet another method of finding out the $\kappa(\alpha)$ dependence. It is computationally efficient and gives more insight into the dynamics of modelled processes.

First, however, let us compare results of two experiments. We will concentrate here on the analysis of $F_Z$, but these simulations will give us $F_Y$ dependencies as well.

Experiment \textbf{I}. The potential energy $E_p$ and force $F_Z$ shown in \ref{fig:myPotential} are for atoms belonging to the upper layer 1, when it is moved out of equilibrium in the X-direction for a certain distance (amplitude $A$), and then released at a time $t=0$, followed by the NVE integration. At the same time a force equal to 0 is imposed on layers 2 and 3 in all 3 directions, and therefore these layers cannot move. 

Experiment \textbf{II}. Layers 1 and 3 are fixed (a force equal to zero is imposed, so atoms there cannot move), and the middle layer 2 is displaced in the X-direction for a distance $A$; next it is released, and the NVE integration is performed.

\begin{figure}[ht]
\centering
\includegraphics[scale=1.0]{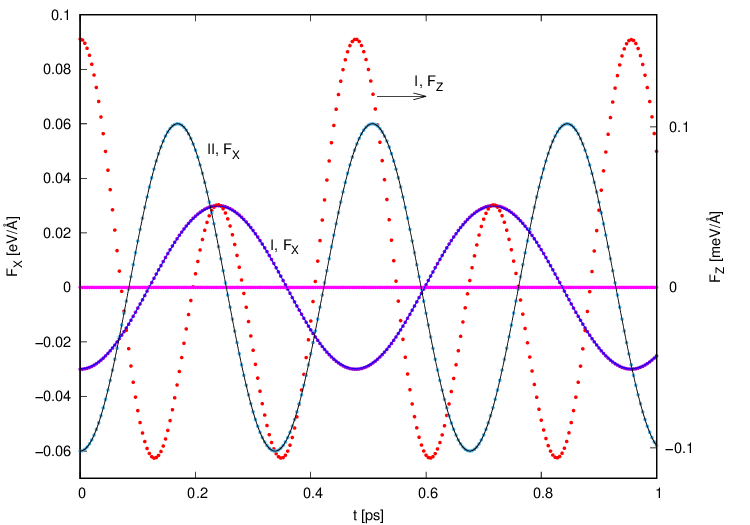}
\caption{
Comparison of forces $F_X(t)$ and $F_Z(t)$ in two simulations, \textbf{I} and \textbf{II}, as described in the text. The amplitude of the initial X displacement is 0.03 {\AA}. Notice the difference of vertical scales for $F_X$ and $F_Z$. In the case of \textbf{II} $F_Z$ is equal to 0.
}
\label{fig:surface00B}
\end{figure}

The results on the time dependence of forces $F_X(t)$ and $F_Z(t)$ are compared for both experiments in figure \ref{fig:surface00B}, and marked there by labels \textbf{I} and \textbf{II}. Notice the scale difference for $F_X$ and $F_Z$. The force in the X-direction is found to be twice as large in case \textbf{II} as in case \textbf{I}. This is due to the fact that in \textbf{II} the layer 2 interacts with 2 layers, while in \textbf{I} it interacts with one layer only, and therefore the effective potential per atom is 2 times larger in \textbf{II}. That also results in the difference of oscillation periods: that one is inversely proportional to the square root of the magnitude of the harmonic potential well. Indeed, in \textbf{II},
$\Omega$=13.144/ps is found, and it is exactly $\sqrt{2}$ times smaller in \textbf{I}.

No forces in the Z- and Y-directions are found in case \textbf{II}. This is the crucial observation of this work: A force in the Z-direction acting on a given layer occurs only when the relative displacement (strictly, a square of a displacement) of the two neigbouring layers on both sides of a given layer in the X-direction are different. In other words, when the displacement of neigbouring layers in the X-direction are the same, there is no force acting on a layer in the Z-direction. In the case of \textbf{I} $F_Z$ is nonzero because layer 1 interacts with only one layer.

We notice also an unexpected dependence of $F_Z(t)$: it looks like a two harmonic contributions are there, one of frequency $\Omega$ and the other one of $2\Omega$ (which is indeed the case, as it will be explained).

When an external pressure is applied in the X-direction, forces acting on layers in the Z-direction must have a contribution from two mechanisms: one is obviously due to a harmonic motion in the Z-direction. That part ought to be given by $- k\cdot \Delta Z$, where $k$ is the same spring constant as the one describing the motion in the X-direction under force in the same direction. The second part, as already argued, is caused by a nonlinear contribution that is in the direction perpendicular to the X axis, and it is defined as $\kappa \cdot \Delta X^2$. Therefore, the entire force $F_Z$ must be given as

\begin{equation} 
	F_Z(t)= - k\cdot \Delta Z(t) + \kappa \cdot \Delta X^2(t).
\label{eq:FZ}
\end{equation} 

\begin{figure}[ht]
\centering
\includegraphics[scale=1.0]{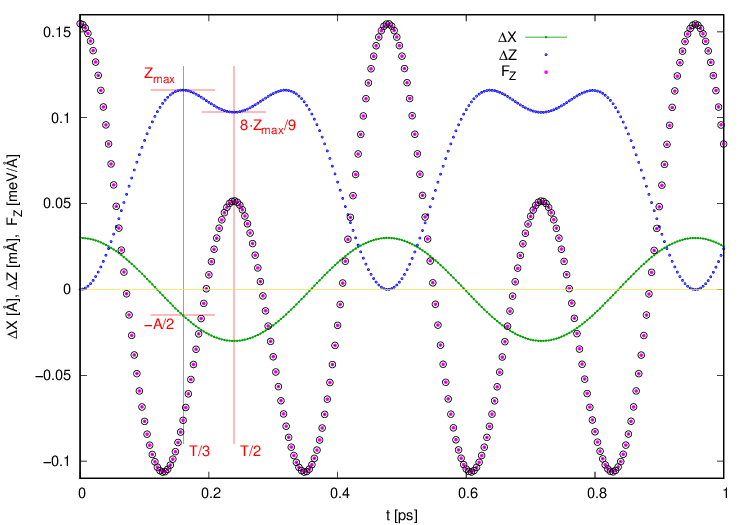}
\caption{
Testing the relation $F_Z(t)= - k\cdot \Delta Z(t) + \kappa \cdot \Delta X^2(t)$.
Notice different vertical scales and units for each quantity. 
The green line shows $\Delta X(t)$, the blue dots are $\Delta Z(t)$, magenta dots is the $F_Z(t)$ dependence. All these are determined from MD simulations. All the data are from the same simulation run as those used in figure \ref{fig:surface00B}.
Large empty black circles are computed by using the above formula, with the value of $\kappa$ determined as described in figure \ref{fig:myPotential}.
}
\label{fig:surface00D}
\end{figure}

That relation is checked in figure \ref{fig:surface00D}. There we compare the value determined from simulations, with a value computed by using \ref{eq:FZ} and the data on $\Delta X(t)$ and $\Delta Z(t)$ taken also from simulations. Hence, this is an illustration of how a mixing of X- and Z-components of the movement must be taken into account for a proper description of dynamics, which is in agreement with the Newtonian mechanics.

\begin{figure}[ht]
\centering
\includegraphics[scale=1.0]{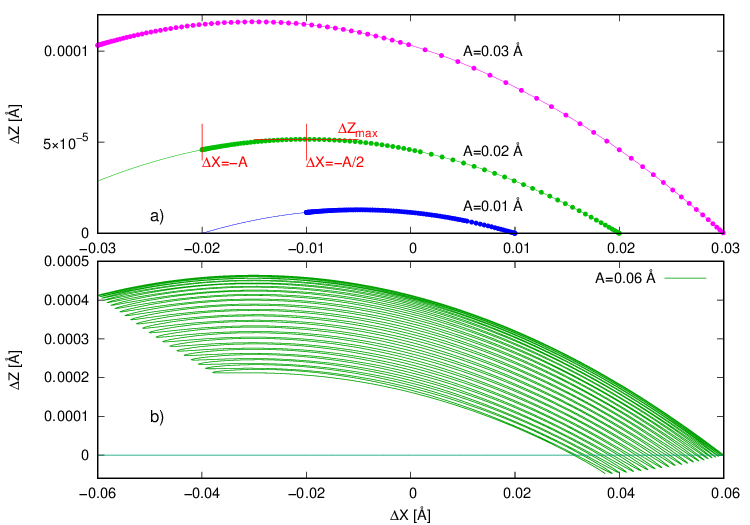}
\caption{X-Z phase-space portraits when $\alpha=0^{\circ}$. In a) the data are for 3 values of the amplitude $A$ of the initial displacement $\Delta X$, 0.01, 0.02 and 0.03 {\AA}. Time is the hidden parameter there. The simulation data are shown as points and fall on parabolic curves (shown by continuous lines).
In b) we show similar data for a larger amplitude $A$=0.06 {\AA} and for a much longer time range, as an illustration of the drift of curves with time. When the movement in the Y-direction is disabled, no any drift with time is observed, and all the data points follow the initial curves determined at the shortest times.
}
\label{fig:surface00G}
\end{figure}

Let us now discuss another type of phase-space diagram, for the case when the upper layer is moved by using the same or similar simulation data as those used to draw figures 
\ref{fig:surface00B} and \ref{fig:surface00D}. In figure \ref{fig:surface00G} a) dependencies of $\Delta Z$ on $\Delta X$ are shown at short times $t \lessapprox T$ (where $T$ is the oscillation period), for 3 values of the initial amplitude $A$, when $\alpha=0^{\circ}$. For all 3 curves, the X-axis scales up linearly with $A$, and the Z-axis scales up as A$^2$.

Since forces in the Z-direction are several orders lower in magnitude than in the X-direction, we may assume as an approximation that the movement in the X-direction is not influenced by $F_Z$. Hence, $\Delta X(t)$ must be given by a simple harmonic function fulfilling the initial boundary condition $\Delta X(t=0)=A$, that is: $\Delta X(t)=A\cdot \cos(\Omega t)$.
Hence, while $X(t=0)=A$, the minimum value of $X$ is at $t=T/2$, i.e., $X(\Omega T/2)=-A$. 
The initial value of $Z$ is $Z(t=0)=0$. The time $t=T/2$ is the return point.
It follows from the data analysis that the maximum value of $\Delta Z(t)$, which is $Z_{max}$, is observed when 
$\Delta X(t)=-A/2$ (compare red horizontal and vertical bars in figures \ref{fig:surface00D} and \ref{fig:surface00G} a). Hence, the maximum of $Z$ occurs at $t=T/3$. At this moment we do not have a derivation of equations of motion allowing us to find an analytical expression on $\Delta Z (\Delta X)$. It is evident, however, that with a large confidence and accuracy, we can approximate that dependence by a parabolic function, and it must be given by the following formula, fulfilling all the required boundary conditions and the observed properties:

\begin{equation} 
	\Delta Z = Z_{max}[1-(4/9)(\Delta X/A+1/2)^2],
\label{eq:DeltaZ}
\end{equation} 

We find from \ref{eq:DeltaZ} the value of $\Delta Z$ at $\Delta X=-A$ as $8Z_{max}/9$.

Actually, there is a small drift of the curves with time (hardly observable at these low amplitudes and short times). It is seen better in \ref{fig:surface00G} b) drawn for a larger amplitude of oscillation, $A$=0.06 {\AA}, and for a much longer time range.

\subsection{Equations of motion in the X-Z plane.}
\label{XZplane}

Since $\Delta X(t)=A\cdot \cos(\Omega t)$, and $F_Z(t)=\kappa (\Delta X(t))^2 - k\Delta Z(t)$, and $\rm{m}\cdot d^2\Delta Z(t)/dt^2 = F_Z$, $k/\rm{m}=\Omega^2$, we are able to write the following differential relation: 

\begin{equation} 
d^2\Delta Z(t)/dt^2 = \kappa/\rm{m} \cdot A^2\cdot \cos^2 (\Omega t) -\Omega^2 \cdot \Delta Z(t),~~~\Delta Z(0)=0,  V_Z(0)=0.
\label{eq:dVZ0}
\end{equation} 

In order to simplify notation, let us replace $\Delta Z(t)$ by $z(t)$. Let us also use the relation $\cos^2(x) = (1+\cos(2x))/2$. Now,  \ref{eq:dVZ0} can be written as

\begin{equation} 
\ddot{z} = \frac{\kappa A^2}{2\rm{m}}\cdot \left(1+\cos(2\Omega t)\right) -\Omega^2 \cdot z.
\label{eq:dVZ1}
\end{equation} 

We recognise that \ref{eq:dVZ1} is an equation of a driven harmonic oscillator (see, e.g., \cite{Thornton}), with a driving force of angular frequency $2\Omega$. One can simply check that \ref{eq:DeltaZ} is a solution of \ref{eq:dVZ1} when we use

\begin{equation} 
	Z_{max} = 3\kappa A^2/(4\rm{m} \Omega^2).
\label{eq:dVZ2}
\end{equation} 

\begin{figure}[ht]
\centering
\includegraphics[scale=1.0]{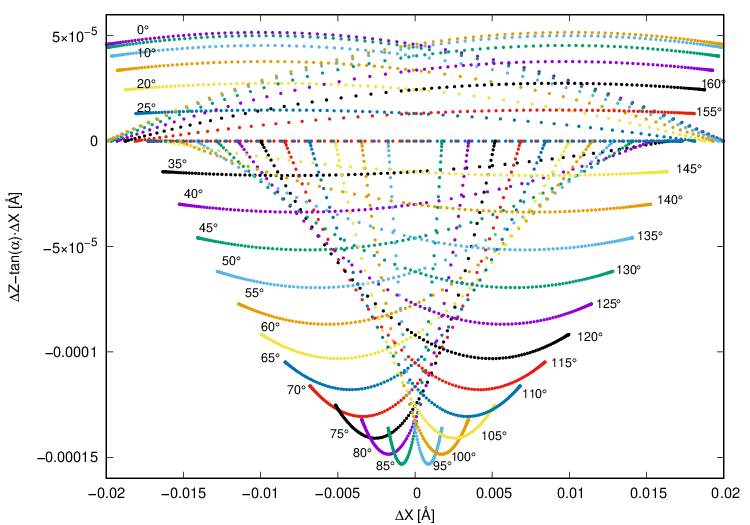}
\caption{Dependence of $\Delta Z - \tan(\alpha)\cdot \Delta X$ on $(\Delta X)$, determined from initial oscillations, when the layer is displaced for 0.02 {\AA} in a direction $\alpha$ with respect to the X-axis. The labels indicate the angle $\alpha$.
}
\label{fig:surface00A}
\end{figure}

Equation \ref{eq:dVZ1} and its solution \ref{eq:DeltaZ} may be treated as approximations only, valid at short times $t \lessapprox T$. In particular, these equations do not provide a description of the mechanism and dynamics behind a small drift of curves with time, as shown in figure \ref{fig:surface00G} b). 

The description used in this section can be extended to analysis of oscillations in arbitrary direction, for showing another, alternate method of finding the angular dependence $\kappa(\alpha)$. Let us restrict ourselves this time to providing the algorithm only of the data analysis, without giving its detailed explanations.

1) For any data obtained from simulations performed at an angle $\alpha$ with respect to the X-axis, we draw the difference between $\Delta Z$ and $\tan(\alpha)\cdot\Delta X$ as a function of $\Delta X$, as shown in figure \ref{fig:surface00A}. 

2) By using the method of least-squares fitting, we determine the parameters of the parabolic dependencies shown there. 

3) The true parameter $Z_{max}$ for any given $\alpha$ is obtained by multiplying the one from step 2) by $\cos(\alpha)$.

Using \ref{eq:dVZ2}, $\kappa$ can be found from $\kappa=4\rm{m}\Omega^2/(3A^2)\cdot Z_{max}$. We do not show the results on $\kappa(\alpha)$ this time because they are practically identical to these shown in figure \ref{fig:surfaceXZ1}. With the mass of 1 atom of Fe, $\rm{m}=9.273\cdot 10^{-26}kg$, $\Omega=13.144/$ps, the initial amplitude of oscillations $A$=0.02 {\AA}, we found out that the best fitting gives us the angular dependence: $\kappa(\alpha)=\kappa_0\cdot \cos(3\alpha)$, with $\kappa_0$=0.17197 eV/{\AA}$^3$, while in figure \ref{fig:surfaceXZ1} $\kappa_0$=0.1717 eV/{\AA}$^3$ was obtained, and the value computed from first principles in subsection \ref{InterlayerPotential} is 0.1720 eV/{\AA}$^3$.

\subsection{Equations of motion in the X-Y plane.}
\label{XYplane}

Equation \ref{eq:Exz2} shows that the harmonic contribution to the potential energy of the Y-component of a displacement is 4 times larger than for directions X and Z. That will result in a 2 times larger frequency of oscillations of layers in the Y-direction, since it scales up as a square root of the harmonic potential curvature.

A full expression on the force in the Y-direction is 
				
\begin{equation} 
F_Y = -\frac{\partial E_p(x,y,z)}{\partial y} =  -\epsilon_0 \cdot 
  \left(4y+\lambda_Y \left( \left(x^2+z^2\right) +2y^2 \right) \right).
	\label{eq:ExderY}
\end{equation}

Based on \ref{eq:ExderY}, when $y=0$, we have a parabolic dependence of force $F_Y$ on a displacement in X- or Z-direction:

\begin{equation} 
F_{Y}(x)= -\epsilon_0 \lambda_Y \cdot \rho^2 \equiv -\kappa_Y \cdot \rho^2,
	\label{eq:ExderY2}
\end{equation}

with $\kappa_Y \approx $0.48653 eV/{\AA}$^3$.

Let us consider the case when an initial displacement has been made in the X-direction for a value of A. Hence, the oscillating movement
occurs in the X-direction with frequency $\Omega$. In this case (we neglect the force contribution which is proportional to $y^2$), the force in the Y-direction is given by:

\begin{equation} 
	F_Y(t)= -4k\cdot y -\kappa_Y \cdot x(t)^2 = -4k\cdot y -\kappa_Y \cdot A^2\cdot \cos^2(\Omega t).
\label{eq:Ydiff}
\end{equation}

\begin{figure}[ht]
\centering
\includegraphics[scale=0.8]{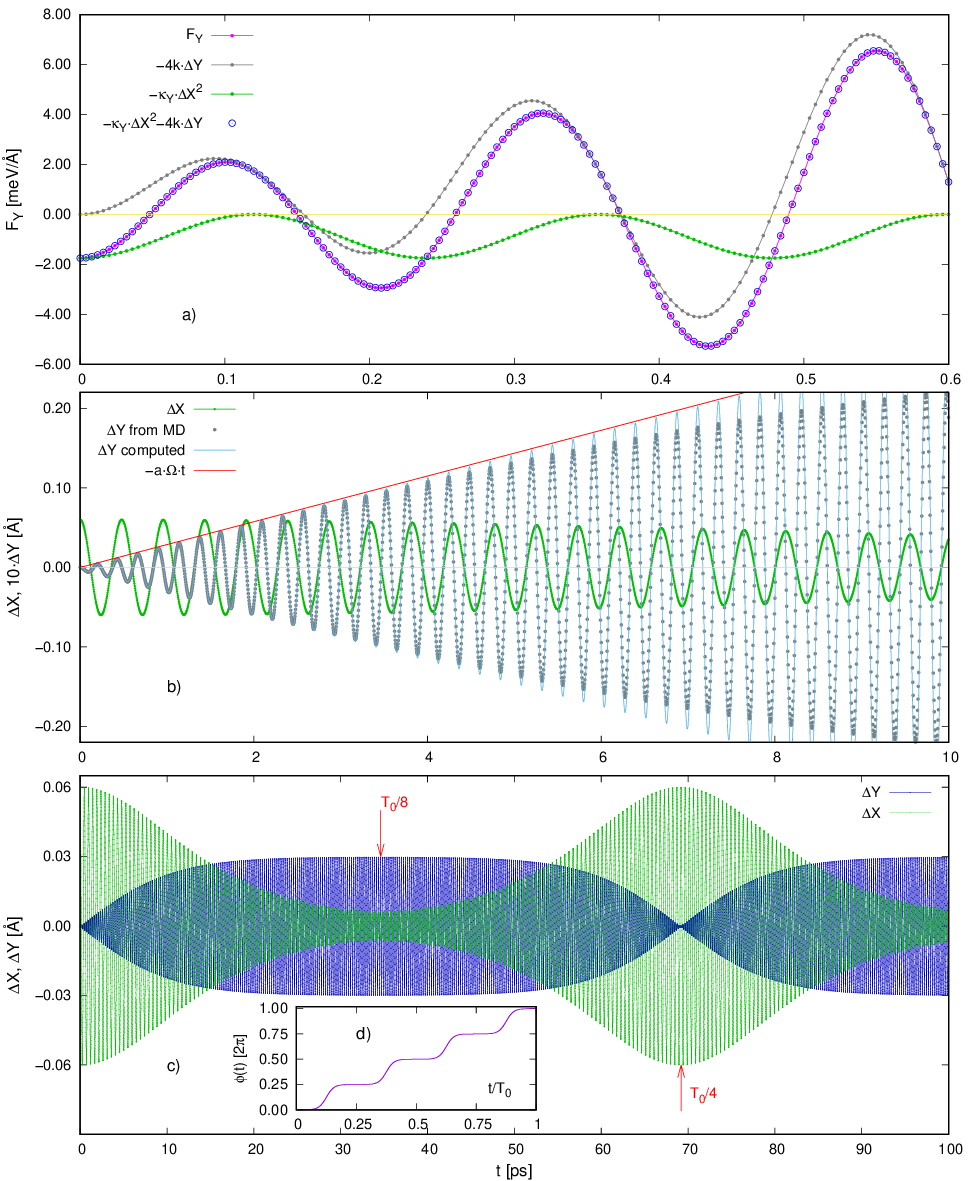}
\caption{The initial amplitude $A$ of X displacement is 0.06 {\AA}.
a) The force component in Y-direction, $F_Y$, is shown as a sum of a few terms. 
b) shows how the amplitude of oscillations in $\Delta X(t)$ and $\Delta Y(t)$ changes with time. 
Notice the vertical scale difference between $\Delta X$ and $\Delta Y$.
c) A long time evolution of $\Delta X$ and $\Delta Y$.
d) presents how the phase of oscillations in $\Delta X(t)$ changes with time.
}
\label{fig:surface00H}
\end{figure}

Hence, we can write equations of motion in a similar way like in subsection \ref{XZplane} in the case of motion in the X-Z plane. By noticing that $4k/\rm{m}=(2\Omega)^2 \equiv \omega^2$, and using the relation $\cos^2x=(1/2)(1+\cos2x)$, we have

\begin{equation} 
\ddot{y}  +\omega^2 \cdot y = -\frac{\kappa_Y A^2}{2\rm{m}}\cdot \left(1+\cos(\omega t)\right).
\label{eq:Ydiff2}
\end{equation} 

The above is again an equation of a driven harmonic oscillator. This time let us use the Laplace transform method to solve it.
We define the transformed $y(t)$ as $y^*={\mathcal L }[{y}](s)$, obtaining \ref{eq:Ydiff2} in the form

\begin{equation} 
s^2 y^* -sy(0)-\dot{y}(0)+  \omega^2 \cdot y^* = a/s +as/(s^2+\omega^2),
\label{eq:Ydiff3}
\end{equation} 

where $a=-\kappa_Y A^2/(2\rm{m})$. Since the initial conditions are $y(0)=0$ and $\dot{y}(0)=0$, we get

\begin{equation} 
y^* = \frac {a}{s(s^2+\omega^2)} +\frac{as}{(s^2+\omega^2)^2}.
\label{eq:Ydiff4}
\end{equation} 

The inverse transform allows us to find the solution, $y(t)={\mathcal L^{-1} }[{y^*(s)}]$:

\begin{equation} 
y(t) = -\frac{\kappa_Y A^2}{8\rm{m} \Omega^2} \cdot \left(1-\cos(2\Omega t)+ \Omega t \cdot \sin(2\Omega t)\right).
\label{eq:Ydiff6}
\end{equation} 

In figure \ref{fig:surface00H} a) we show that the initial force $F_Y$ is given as a sum of a force caused by the usual harmonic movement in that direction, which is $4k\cdot \Delta Y$, and a part caused by a displacement in the X-direction, which is $\kappa_Y \cdot (\Delta X)^2$, with $\kappa_Y$ value from equation \ref{eq:ExderY2}. Hence, we demonstrate the validity of the relation $F_Y(t)= -4k\cdot \Delta Y -\kappa_Y \cdot (\Delta X)^2$. The contribution to force from $y^2$ can be neglected in this case.

Equation \ref{eq:Ydiff6} may be treated as a good approximation for very short times only, as long as $y(t) \ll x(t)$. This is well visible in figure \ref{fig:surface00H} b). The dots labelled as "$\Delta Y$ from MD" are simulation results, while the continuous curve labelled "$\Delta Y$ computed" (with corresponding red line showing its slope $a$) is calculated by using equation \ref{eq:Ydiff6} with $\kappa_Y A^2/(8\rm{m} \Omega^2) \equiv$ $a=2.186~10^{-5}$ {\AA}. At the same time, when $\kappa_Y$ is used from equation \ref{eq:ExderY2}, we get $a=2.433~10^{-5}$ {\AA}.

A deeper understanding of the phenomena is gained from analysis of a longer time evolution of oscillations, as shown in \ref{fig:surface00H} c). We observe a long-term periodicity in the amplitude of $x(t)$ and $y(t)$. The arrow labelled as $T_0/4$ (at 69.2 ps) is actually not the full period of these oscillations but a 1/4th of the full period. Since we are able to determine $\Omega$ with a high accuracy, we could find out (by a numerical analysis of the data) how the phase of $x(t)$ changes with time, when that phase is defined as in the relation $x(t)=A(t)\cdot \cos(\Omega t + \Phi(t))$. The dependence $\Phi(t/T_0)$ is shown in \ref{fig:surface00H} d). Hence, a full change of phase for $2\pi$ occurs at $t=T_0$, while at $t=T_0/8$, etc, the phase changes for $\pi/2$. The fastest changes take place at times close to integer multiples of $T_0/8$. 

Changes of phase in $y(t)$ are somewhat different. The phase is nearly time independent in broad time ranges, with a very narrow time range (of around $2\pi/\Omega$) near multiples of $t=T_0/4$, where an abrupt change is observed for $\pi$. Hence, in the case of $y(t)$ a phase change for $2\pi$ occurs at $t=T_0/2$.

The above is not necessarily the only possible schema of phase changes.

The period $T_0$ is found to be inversely proportional to the initial amplitude $A$ of oscillation in $x(t)$, when $A$ is lower than around 0.2 {\AA}. That leads to an appearent square proportionality of the initial slope of the growth of amplitude of $y(t)$ on $A$, as given by equation \ref{eq:Ydiff6}.

\begin{figure}[ht]
\centering
\includegraphics[scale=1.0]{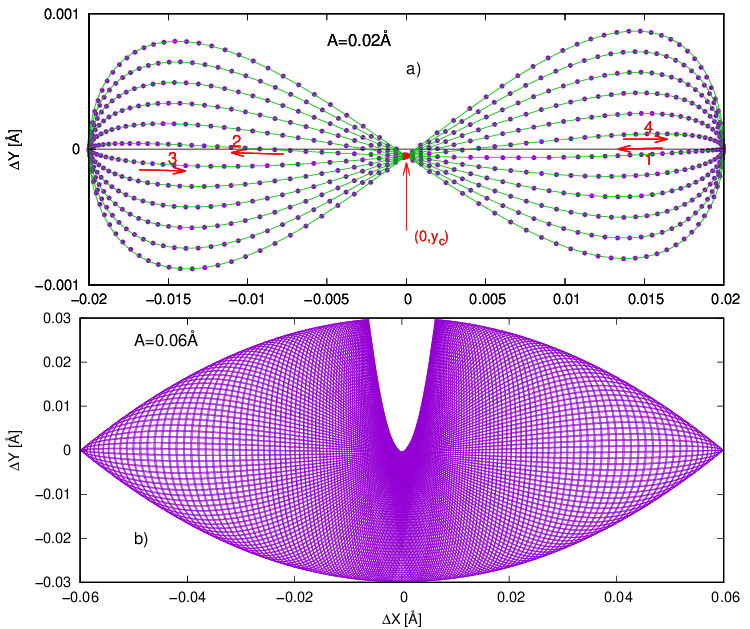}
\caption{Trajectories of particles in the Z-Y phase space.
a) The first 6 full loops when $A$=0.02 {\AA}.
b) Long-time oscillations when $A$=0.06 {\AA}. 
}
\label{fig:surface00J}
\end{figure}

These observations indicate that the total energy of the system does not change with time. That is confirmed by analysis of velocities (which are related to the kinetic energy $E_k$). We observe that the initial potential energy accumulated in the system due to a displacement of X for $A$ converts fully into $E_k$ at $t=T_0/8$ in the Y-mode of oscillations, and then it is transferred back to the X-mode at $t=T_0/4$. Additionally, we were able to confirm that the total energy (which is the sum of potential energy and kinetic energy) does not change with time, with a high relative accuracy up to 6 significant digits, by drawing the total energy as a function of time.

\subsection{Trajectories in the X-Y plane.}
\label{Chaos}

Figure \ref{fig:surface00J} a) illustrates an X-Y trajectory of the uppermost layer under the driving force of an angular frequency $\Omega$ in the X-direction of amplitude $A$=0.02 {\AA}. The starting point at t=0 is at (0.02,0). The first 6 periods of oscillations are shown. The arrows labelled 1-4 indicate directions of the movement of the layer.

Dots are from simulations (the same data are used as in \ref{fig:surface00H} b), while lines are drawn by using a parametric set of variables $y(t)$ as given by \ref{eq:Ydiff6}, with $x(t)=A\cdot\cos(\Omega t)$.

\ref{fig:surface00J} b) shows similar results for a much longer time, exceeding $T_0$, when the initial amplitude $A$ is 0.06 {\AA}. 

The point of crossing of loops in \ref{fig:surface00J} a), marked as 
$(0, y_{c})$, can be determined from equation \ref{eq:Ydiff6}. It occurs when both the sine function equals to 0 and the cosine is -1 in \ref{eq:Ydiff6}. Therefore, $y_{c} = -(\kappa_Y A^2)/(4\rm{m} \Omega^2)$. However, this is an approximation only. At longer times that position changes slightly.

\section{Discussion.}
\label{discussion}

\subsection{A connection with Christoffel equations.}

In the theory of elastic waves in crystals, Christoffel equations \cite{Fedorov,Levy,Ting1,Ting2,Jaeken} relate the static elastic constants with possible directions of plane wave propagation and their polarisation (velocity vectors of the displacement of atoms) \cite{Lynch}. For an FCC crystal, there are only three independent elastic constants \cite{Levy}. We determined values of stiffness parameters by using the method of Sprik et al \cite{Sprik} available in the source code distribution of \texttt{lammps} \cite{LAMMPS} through a package written by Aidan Thompson. The values are: $C_{11}$=180 GPa, $C_{12}$=90 GPa, and $C_{44}$=90 GPa. 

In the case of waves traveling in the [1,1,1] direction, values of corresponding sound velocities are ($\rho$ is density): 
$c_L=\left((C_{11}+2C_{12})+4C_{44})/3\rho\right)^{1/2}$ (the longitudinal one) and $c_T=\left((C_{11}-C_{12}+C_{44})/3\rho\right)^{1/2}$ (two degenerate transverse components; compare this with Levy \cite{Levy}). Hence, we compute: $c_L$=54.031{\AA}/ps and $c_T$=27.016 {\AA}/ps. $c_T$ can be found easily from data like these in figures \ref{fig:XYZL00A2}-\ref{fig:XYZL00E2}, while $c_L$ from similar simulations, when a Heaviside pressure is applied along the Y-direction. Values of the sound speed components found from MD simulations are in very good agreement with these computed above when simulations are performed in the limit of very small applied pressure (let say below 10 MPa).

\begin{figure}[ht]
\centering
\includegraphics[scale=1.0]{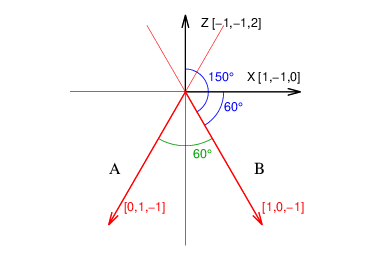}
\caption{Eigenvectors of solutions to Christoffel equations for transverse waves travelling in the [1,1,1] direction.
}\label{fig:XYZLT}
\end{figure}

Eigenvectors of Christoffel equations determine polarisation of velocities. 
The longitudinal velocity is in the [1,1,1] direction. The other two, degenerate transversal velocities are in directions [1, 0, $\bar{1}$] and [0, 1, $\bar{1}$] (figure \ref{fig:XYZLT}). Some authors suggest (e.g., Levy \cite{Levy}, p. 18.) that it does not matter what is the direction of oscillations of the transverse wave, as long as its displacement component lies within the (1,1,1) plane. That seems a natural assumption since Christoffel equations are linear ones and therefore solutions in any direction within that plane can be constructed as a linear combination of two transverse components in directions [1, 0, $\bar{1}$] and [0, 1, $\bar{1}$]. 
In our case, as it follows from the angular dependence of equipotential diagrams (\ref{fig:myPotentialA}) as well as from figure \ref{fig:surfaceXZ1}, angles of $\ang{30}$ ($\pm$ integer multiples of $\ang{60}$), have a special meaning: the contribution to the particle motion that is perpendicular to the applied force disappears. These are directions that are perpendicular to one of the eigenvectors that are solutions to Christoffel equations (figure \ref{fig:XYZLT}). 
At first sight one might think that at these special angles we are in the limit of a linear problem, since coupling between the movement in the $\hat{\rho}$ and $\hat{\alpha}$ directions disappears. However, this is a wrong assumption: we do observe at these angles (not shown) a strong dependence of MD simulation results on the amplitude of the applied force. 

We conclude that Christoffel equations may be treated as an approximation only valid in the limit of very small pressure applied.

\subsection{H{\'e}non-Heinen potential}

Limiting the motion to the X-Z plane in equation \ref{eq:Final00xyz} is, in \texttt{lammps}, a matter of imposing proper conditions on forces allowed. In that case, with the $y$-component of force neglected, the potential is reduced to  

\begin{equation}
E_p/\epsilon_0   = \frac{1}{2} \left(x^2+z^2\right) 
	+\lambda \cdot\left(x^2 z - \frac{z^3}{3}\right).
	\label{eq:HH}
\end{equation} 

The potential given by \ref{eq:HH} has been first introduced by H{\'e}non and Heinen \cite{Henon}, \cite{Henon2} as a model of the potential reproducing dynamics of stars in the centre of the galaxy. It contributed significantly to the development of the theory of chaos. At present times the dynamics resulting from movement of bodies in that potential is still investigated intensively \cite{Zhao}, \cite{Chandra},\cite{Kostov}, \cite{Zotos}, \cite{Sawsan}, \cite{Costin}. However, with some exceptions, the subject did not find a recognition deserving it in fields distant from astrophysics or mathematical physics. 

Equation \ref{eq:HH} is usually studied assuming $\lambda=1$ since in that case its solutions become dependent on one parameter only, the (dimensionless) energy of a particle at the initial stage of modelling. Hence, for instance, it is observed (when $\lambda=1$) that for energies lower than 1/12 regular nonergodic dynamics is found, with closed loops in phase space (though these trajectories might be very complex), while at larger energies chaotic motion emerges (with apparently no closed trajectories) \cite{Henon}, \cite{Zhao}, \cite{Zotos}. 
For energies larger than 1/6 the system will escape the bound region at a finite time, with probability dependent on energy \cite{Zhao}, \cite{Zotos}, \cite{Costin}.

Equation \ref{eq:HH} is scalable in the sense that changing the scale of coordinates does not change the equation itself, but the scale of energy and $\lambda$ change.

In our case $\lambda$=0.1720 eV/{\AA}$^3$. and it is related only to the distance between atoms, $\lambda\sim 1/r_0$. Therefore we would need to decrease $r_0$ by a factor 1/0.1720 in order to reach the value of $\lambda=1$. We do not know about FCC crystals that have such a short distance between atoms. On the other hand, we report in section \ref{Other_comments} that in some materials the apparent value of 
$\lambda$, as determined from MD simulations, is much larger.

Our 3D version of \ref{eq:HH}, given by \ref{eq:Final00xyz} (which is scalable in the same way as equation \ref{eq:HH}), has not been mathematically studied so far and it likely offers a broad field for exploration as well. In particular, we do expect complex dynamics there due to, for instance, (chaotic) transitions of particles between different crystallographic cells. Similar escape transitions have been recently modelled on the surface of graphene and silicene near the melting point \cite{Hassan}. An intriguing question is what is the possible role of escapes in the context of crystallographic phase transformations \cite{Jun}.

The interatomic potential we are using in \texttt{lammps} is, however, not suitable for similar simulations.
It has been intentionally designed to limit interaction between atoms to the nearest neighbours only by using a cutoff (at 3 {\AA} in our case).

\subsection{Analytical solutions of equations of motion in a CSM model.}

Exact analytical solutions of equations of motion in a 1D CSM model \cite{Stretched}, as given by \ref{eq:uH}-\ref{eq:fH},  are derived from a set of linear difference equations:

\begin{equation}
	{\ddot{x}}_{n} + 2{{\rm{\Omega }}}^{2}{x}_{n} 
			- {{\rm{\Omega }}}^{2}{x}_{n+1}
			-{{\rm{\Omega }}}^{2}{x}_{n-1}=0,
\label{eq:XZ0}
\end{equation}

with the boundary condition:

\begin{eqnarray}&&{\ddot{x}}_{1}+{{\rm{\Omega }}}^{2}({x}_{1}-{x}_{2})={F}_{1}(t)/\rm{m},
\label{eq:XZ1}
\end{eqnarray}

where $x_n$ is the displacement of the n-th layer, $F_1(t)$ is the time-dependent force acting on the layer 1, and $\Omega$ is the angular frequency of a single mass $\rm{m}$ with spring constant $k$ of harmonic interaction between two masses, $\Omega=\sqrt{k/\rm{m}}$.

Let us consider the 2D case of the H{\'e}non-Heinen potential \ref{eq:HH}. Now we would need to construct a pair of difference equations, each of them nonlinear, for every n-th layer, from equations of motion following from \ref{eq:HH}:

\begin{eqnarray*}
	\ddot{x} &=  -{\Omega }^2 x - 2 \epsilon_0 \lambda /\rm{m} \cdot x z\\
	\ddot{z} &=  -{\Omega }^2 z - \epsilon_0 \lambda /\rm{m} \cdot (x^2- z^2)
\label{eq:HHM}
\end{eqnarray*}

It is very unlikely that equations will have a closed-form exact analytical solution. 
It might be possible to compute approximate numerical solutions by using an iterative method for a set of n layers (with 4 equations for each layer, when counting initial boundary conditions). Is it not better, however, to use in that case \texttt{lammps} as a solver?

\subsection{Other comments.}
\label{Other_comments}

The resulting dynamic is a complex one; this is a field requiring an additional systematic exploration.

The presented model implies the existence of a displacement of layers in the Z-direction (and the Y-direction as well) when a force is applied in the X-direction. This effect is a dynamic one, and it is inconsistent with the static theory. However, it becomes significant only at large applied forces: since it is proportional to the square of pressure, it may have been overlooked in experiments. 

It might be worthy of considering an experimental verification of the discussed phenomena on crystals and/or on artificial acoustic metamaterials. We would expect that observations need to be conducted at microscopic scales and very short times, perhaps with the use of laser light as a mechanism to excite acoustic waves in crystals, as for instance studied by the FEM method \cite{He}.

While the simulation setup used in the experiments described here has been simplified as much as possible (harmonic interlayer potential is used with an ideal monoatomic crystal structure), the results of simulations performed by us on more realistic structures (several kinds of FCC compounds, with various types of atoms and atomistic potentials, such as steel 310S with EAM interatomic potentials (Artur and Bonny) and medium entropy alloys CoNiCr and CoNiV, with EAM and MEAM potentials, are qualitatively the same. The $\cos(3\alpha)$ dependence is found in all of them. We observed, though, that there is a large spread of the values of $\kappa_0$ between these compounds (from 0.087 eV/{\AA}$^3$ for CoNiCr with EAM potential to 3.63 eV/{\AA}$^3$ for steel with Bonny potential), which, probably, ought to be attributed to differences in the chemical inhomogenity of compounds, leading to large deviations from perfect symmetries of atom arrangement. We observe also lattice displacements and large forces in the Z-direction, of a similar, oscillating in time/position, when the front of a  stress wave propagating in the Y-direction hits a layer. The effect raises a question, in particular, about its role in the dynamics of the movement of dislocations.

\section{Conclusions.}
\label{Conclusions}

A dynamical effect has been observed in MD simulations and explained by a simple mechanism, supported by an analytical model. It is caused by broken symmetry of interaction between neighbouring crystallographic layers by propagating stress waves. The effect is not due to the displacement between the nearest layers alone but due to the difference between displacements of layers on both sides of any given layer. A travelling stress wave causes that difference. Hence, it is not a surface effect but a bulk one. We describe it with an example of a particular crystallographic orientation of an FCC crystal. However, it ought to be present also in some other simulation setups, when there is a difference between atomic configurations on two sides of a layer and stress waves travel in a perpendicular direction to layers.

One of the consequences is, in our case, a dynamically broken axial symmetry in the (111) plane, leading to a threefold symmetry instead, as demonstrated by simulations of angular dependence of forces, as well as a nonzero shear displacement in the Z-direction when force is applied in the X-direction. We expect that such a displacement ought to be observable in experiments performed under appropriate conditions. These effects are proportional to the square of applied forces, and therefore their amplitude tends to zero when forces are small. 

The found connection between the derived analytical potential energy of a layer and the H{\'e}non-Heinen potential opens access to a broad field of mathematical exploration and computer modelling.

\section*{References}
\label{References}

\bibliographystyle{iopart-num}
        \bibliography{Z2S}
\label{Bibliography}

\end{document}